# Mn(II)-doped 2D perovskite for light emitting devices

Daniele Cortecchia, Wojciech Mróz, Stefanie Neutzner, Tetiana Borzda, Giulia Folpini, Annamaria Petrozza*

*Istituto Italiano di Tecnologia, Milan, 20133, Italy*
*Corresponding Author: Annamaria.Petrozza@iit.it*

**SUMMARY**

Low dimensional perovskites are considered good candidates for light emitting applications given the high exciton binding energy which should in principle improve the radiative recombination efficiency. Yet, single-layered two-dimensional (2D) perovskite films are strongly limited by trap-assisted recombination and suffer from low luminescence yields, hampering their application in electroluminescence devices. Here, we use *ad hoc* synthetic and defect engineering strategies to overcome such issue. We employ metallic doping to controllably introduce luminescent impurities in a matrix made of 2D perovskite $NMA_2PbX_4$ based on the cation NMA = 1-naphtylmethylammonium. By means of temperature-dependent and time-resolved spectroscopy we demonstrate efficient energy transfer to $Mn^{2+}$ centres. Such process avoids the funnelling of the photo-excited species in inefficient recombination channels represented by intra-gap trap states and enhances photoluminescence, with quantum yield surpassing 20% in doped films. Eventually, we embody Mn-doped $NMA_2PbBr_4$ in a light emitting diode architecture and show, for the first time, electroluminescence from the $Mn^{2+}:^4T_1 \rightarrow ^6A_1$ transition. This proof-of-concept demonstration shows the potential of doping in layered perovskites and prompt for the study of a wider range of host/guest structures.

**INTRODUCTION**

In the last years, metal halide perovskites have been attracting increasing interest for light emitting applications such as light-emitting diodes (LEDs) and field-effect transistors (FETs), lasing and scintillators.[1-7] Three-dimensional (3D) perovskites have emerged as the best candidate for high-efficiency solar cells,[8] and they have been demonstrated recently also in efficient LEDs.[9-11] Nevertheless, the research community keeps exploring alternative structures to improve the efficiency of radiative recombination.[12,13] The possibility to sensibly increase the exciton biding energy (up to few hundreds meV) by exploiting spatial and dielectric confinement in low-dimensional structures formed by the alternation of organic and inorganic sheets has revived the interest in two-dimensional (2D) perovskites.[14,15] Impressive results have been achieved with multidimensional Ruddlesden-Popper perovskites, where the creation of perovskite films with self-organized heterogeneous phases is exploited to realize a cascade energy transfer from 2D perovskites with large band gap to 3D perovskite domains with narrower band gap, achieving photoluminescence quantum yield (PLQY) values up to 60% and external quantum efficiencies (EQE) beyond 14% in LEDs.[16-22] However, with the exception of photoluminescence measured on exfoliated perovskite single crystals,[23] most of the single-layered 2D perovskites reported so far suffer from very low PLQY <3% when processed in form of thin films, which have frustrated the fabrication of efficient LEDs and their room temperature operability.[20,24-30] Although the origin of the inefficient radiative recombination in these single-layered 2D structures is not yet fully understood, it is likely to stem from fast exciton quenching at room temperature and their high trap density,[31-33] which is consistent with the observation of ultrabroadband largely Stokes shifted luminescence in several compounds of this family.[24,34] Careful synthetic design and defect engineering might represent the key to avoid rapid non-radiative quenching and improve the luminescence yield and tunability of 2D perovskites. Two strategies can be identified, involving the use of photoactive organic cations or the addition of selected heteroatoms to the perovskite lattice. In the first case, energy transfer to functional templating cations (such as quaterthiophene derivatives) has been exploited to improve the light emitting properties of layered perovskites, in some cases with successful integration in electroluminescent devices.[35-37] On the other hand, doping with transition metals ($Mn^{2+}$) and rare earth metals (especially $Yb^{3+}$ and $Ce^{3+}$) have been recently investigated in 3D perovskites nanocrystals such as $CsPbBr_xCl_{3-x}$, where the luminescence from the energy levels of the dopant led to a substantial PLQY enhancement.[38-49] For example, PLQYs of Mn-doped $CsPbCl_3$ nanocrystals typically reaches 30-40%;[50] however, such high efficiencies are usually reported in colloidal dispersions, while it is challenging to preserve them in solid-state, due to the severe luminescence quenching encountered in $CsPbCl_3$ nanocrystals when deposited as thin films.[23,51] Thanks to the wider band gap compared to their 3D homologues, 2D perovskites can serve as excellent host materials to achieve efficient energy transfer to guests dopants. Among 2D perovskites, successful doping with PL enhancement comparable to 3D perovskite nanocrystals has been reported only in $BA_2PbBr_4$ (BA=butylammonium),[52] however the mechanisms underlying the luminesce enhancement of these systems are yet to be clarified. Moreover, to date, there are no reports about their actual integration in electroluminescent devices.

In this work, we investigate the use of a photoactive organic cation and an inorganic dopant to overcome the fast exciton quenching in single-layered 2D perovskites by using $(NMA)_2PbX_4$ (where NMA = 1-naphtylmethylammonium) as a host matrix for manganese doping. NMA is a molecule with energy of the triplet state comparable to the $Mn^{2+}$: $^4T_1 \rightarrow {}^6A_1$ transition characteristic of the inorganic dopant, and are both ideally placed to allow an active role in the luminescent properties of the material.[53-56] By means of temperature dependent and time-resolved spectroscopy, we show that energy transfer from the perovskite's $PbBr_4^{2-}$ inorganic layers can occur to both the NMA cation and the $Mn^{2+}$ guest ions with consequent spectral tuning of the perovskite emission. However, only in the case of metallic doping highly efficient energy transfer is achieved, allowing to overcome the exciton trapping process and leading to luminescent enhancement with PLQY>20% in spin-coated bulk films. To test the effectiveness of the doping strategy for light-emitting devices, for the first time we employed $Mn^{2+}$-doped $NMA_2PbBr_4$ as an active material achieving orange electroluminescence in LEDs operable at room temperature. This proof-of-concept demonstration paves the way to new synthetic strategies for improving the luminescence properties and spectral tunability of low-dimensional perovskites.

## RESULTS AND DISCUSSIONS

Films of $NMA_2PbBr_4$ were spin-coated from DMF solutions by dissolving the organic salt (NMA)Br and the inorganic precursor $PbBr_2$ in 2:1 stoichiometric ratio, while for the doped material $MnBr_2$ was added to the spin coating solution. Doped samples are indicated here with the notation $NMA_2PbBr_4$:Mn(x), where x indicates the molar ratio of manganese with respect to lead in the starting solution. The formation of the perovskite $NMA_2PbBr_4$ was confirmed by X-ray diffraction performed on both powders and thin films (Fig. 1A and S1): as typically observed with such 2D perovskites, in the case of the spin-coated film only some of the diffraction peaks are visible (4.73°, 9.37°, 14.05°) due to the strong preferential orientation of the perovskite on the substrate (Fig. 1A). To study the effects of Mn inclusion on the perovskite structure, we investigated $NMA_2PbBr_4$:Mn(x) over the compositional range x=0-30% mol (Fig. 1A). Upon manganese addition, the host structure is perfectly retained and the absence of additional diffraction peaks excludes the phase segregation of a Mn-based perovskite (for example $NMA_2MnBr_4$, Fig. 1A). A gradual change in the relative intensities of the host diffraction peaks is observed, which can be due either to a change in the scattering factors after doping or to a slight difference in the preferential orientation induced by the presence of Mn. Previous works have demonstrated that manganese doping in 3D lead perovskite nanocrystals proceeds with partial substitution of the octahedrally coordinated $Pb^{2+}$ with the isovalent ion $Mn^{2+}$. Given the bigger ionic size of $Pb^{2+}$ (133 pm) compared to $Mn^{2+}$ (81 pm) lattice contraction is expected, leading to the shift of diffraction peaks to higher angles.[39,42,50] However, in this system, we did not observe a significative change in peaks position or width: this suggests that the crystallization process allows only a small fraction of $Mn^{2+}$ to enter the $NMA_2PbBr_4$ lattice, without therefore resulting in sensible structural change or strain, regardless of the Mn concentration in the precursor solution.

Despite the minor structural changes, Mn addition has a clear impact on the optical properties. The $NMA_2PbBr_4$ absorption spectrum shows the sharp excitonic absorption typical of 2D perovskites peaked at 385 nm, further confirming the formation of the layered perovskite structure in which the exciton binding energy is strongly enhanced due to quantum and dielectric confinement effects. With Mn addition a clear, albeit small, 3 nm blue-shifts and 3 nm bandwidth broadening of the excitonic peak is observed (Fig. 1B and S2). A similar effect involving the blue-shift of the band-edge absorption was also reported in Mn-doped $CsPbCl_3$ and CdSe nanocrystals, and can be attributed to the substitutional alloying of $Mn^{2+}$ in the $PbBr_4^{2-}$ framework which partially disrupt the orbital hybridization of the inorganic skeleton of the host.[42,57,58] The change in the absorption spectrum is extremely similar for all the doped films, irrespective of the Mn/Pb ratio, indicating that the lattice reaches saturation already at the lowest Mn concentration in the precursor solution.

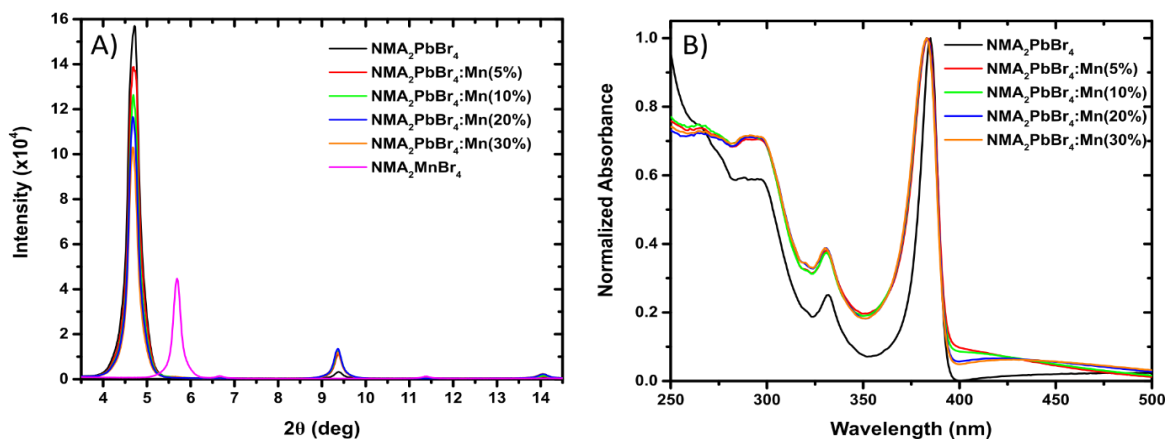

**Figure 1. Impact of Mn addition to NMA$_2$PbBr$_4$.** A) X-ray diffraction (XRD) patterns and B) normalized absorption spectra of NMA$_2$PbBr$_4$:Mn(x), with x = 0-30%.

While having minor effects on the perovskite absorption (see also temperature dependent absorption in Fig. S3), manganese inclusion heavily affects the luminescence properties of the material. Upon photoexcitation of NMA$_2$PbBr$_4$, the main free-excitonic emission is observed at 390 nm (3.18 eV), jointly with a broad luminescence tail in the range 400-500 nm, likely arising from intra-gap defect states with a wide energy distribution (Fig. 2a). Additionally, the broad triplet emission from the NMA$^+$ organic cations appears as a weaker broad band peaked at 550 nm. On the other hand, in the doped films the free excitonic emission is almost completely quenched, while a bright orange emission (peaked at 600 nm) originating from the $^4T_1 \rightarrow {}^6A_1$ transition in the 3d shell of the Mn$^{2+}$ dominates the spectrum (Fig. 2b; see Fig. S8 for the detail of excitonic emission). We note that the quenching of the free-excitonic luminescence in favour of the Mn$^{2+}$ emission is more marked than in Mn-doped CsPbCl$_{x-3}$Br$_3$ nanocrystals,[42,50] indicating a more efficient energy transfer to the guest ions in the case of the 2D perovskite NMA$_2$PbBr$_4$. The change in emission wavelength is accompanied by a remarkable PLQY enhancement from less than 1% (pristine perovskite) up to 22% in doped perovskite thin films. Although we reached the maximum PLQY value at the Mn$^{2+}$ concentration x=10% mol, we note that the PLQY is barely affected by the Mn$^{2+}$ concentration in the precursor solution, and remains nearly constant (PLQY~20%) within the range x=2%-30% mol Mn$^{2+}$ (Fig. S4). We have also tested the iodide-based perovskite NMA$_2$PbI$_4$. Here the NMA phosphorescence is not visible and Mn$^{2+}$ doping is much less effective, since manganese emission only appears as a weak shoulder of the main free-excitonic emission (Fig. 2C,D). This observation is consistent with previous reports on CsPbX$_3$ nanocrystals, where the intensity of Mn luminescence decreases in the series Cl->Br->I following the reduction in thermodynamic driving force for the energy transfer to the doping centres. This is due to the perovskite band gap narrowing and the consequent unfavourable relative position of the energy levels of host and dopant.[42,50]

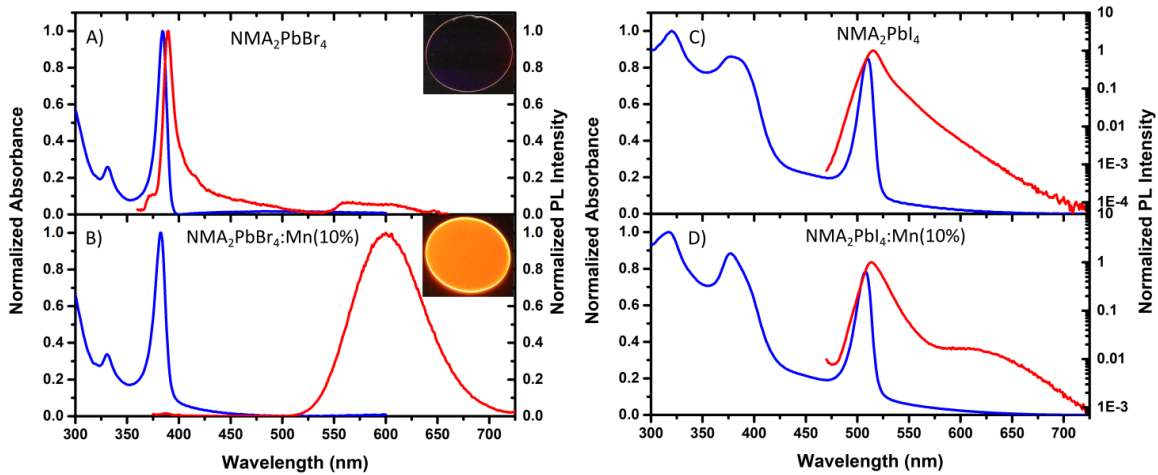

**Figure 2. Doping effects on perovskite's photoluminescence.** Normalized absorption (blue) and photoluminescence spectra (red) of A) $NMA_2PbBr_4$ - $\lambda_{exc}$=340nm; B) $NMA_2PbBr_4$:Mn(10%) - $\lambda_{exc}$=365nm; C) $NMA_2PbI_4$ - $\lambda_{exc}$=450nm; D) $NMA_2PbI_4$:Mn(10%) - $\lambda_{exc}$=450nm. The insets in (A,B) show the photoluminescence of non-doped and doped perovskite films under UV lamp excitation (366 nm).

Differently from the standard alkylammonium cations such as butylammonium, 1-naphtylmethylammonium (NMA) is a functional organic cation which enables an active role in the optoelectronic properties of the perovskite.[12,36,37,54] In fact, the organic salt precursor (NMA)Br at room temperature shows intense fluorescence peaked at 350 nm jointly to a broad, weak phosphorescence in the range 550-650 nm (Fig. S5). When NMA is embedded in the perovskite lattice, the structural rigidification jointly with the heavy atom-effect (the presence of Pb and Br relaxes the spin selection rule for the formally forbidden $T_1 \rightarrow S_0$ transition) considerably enhance its phosphorescence, which becomes clearly visible (Fig. 3A).[53,55] At low temperature the NMA triplet emission is further resolved, showing the vibronic progression of the phosphorescence with the main two transitions peaked at 560 nm and 605 nm (see also the normalized PL spectrum in Fig. S6). The photoexcitation map (Fig. 3B) shows an excellent match between the triplet emission band and the perovskite absorption profile, indicating transfer from the excitons in the $PbBr_4^{2-}$ layers to the triplet of the organic layers. However, a closer look at the spectrum highlights a strong competition between the energy transfer towards the triplet state and the exciton trapping processes, leading to a broad PL tail in the range 400-500 nm. On the other hand, in $NMA_2PbBr_4$:Mn(10%) the Mn: $^4T_1 \rightarrow {}^6A_1$ luminescence undergoes a continuous redshift and narrowing when decreasing temperature (full width at half maximum, FWHM from 90 nm to 55 nm in the range 360K – 77K; Fig. 3C and Fig. S7); this is expected for substitutional doping where $Mn^{2+}$ is in pseudo-octahedral coordination and lattice contraction at lower temperature results in the enhancement of crystal-field strength, thus decreasing the energy of the $^4T_1 \rightarrow {}^6A_1$ transition.[51,59] The excitation spectrum of the transition based on the Mn 3d orbitals perfectly matches the perovskite absorption, indicating an excellent energy transfer from the perovskite host structure to the guest $Mn^{2+}$ ions. Thanks to band narrowing and red-shift, at temperatures below 120K it is possible to disentangle the NMA phosphorescence from underneath the main broad manganese emission, appearing as a weaker shoulder at 560 nm. This reveals that both radiative channels remain active even in the doped system, with the Mn luminescence being the dominant component.

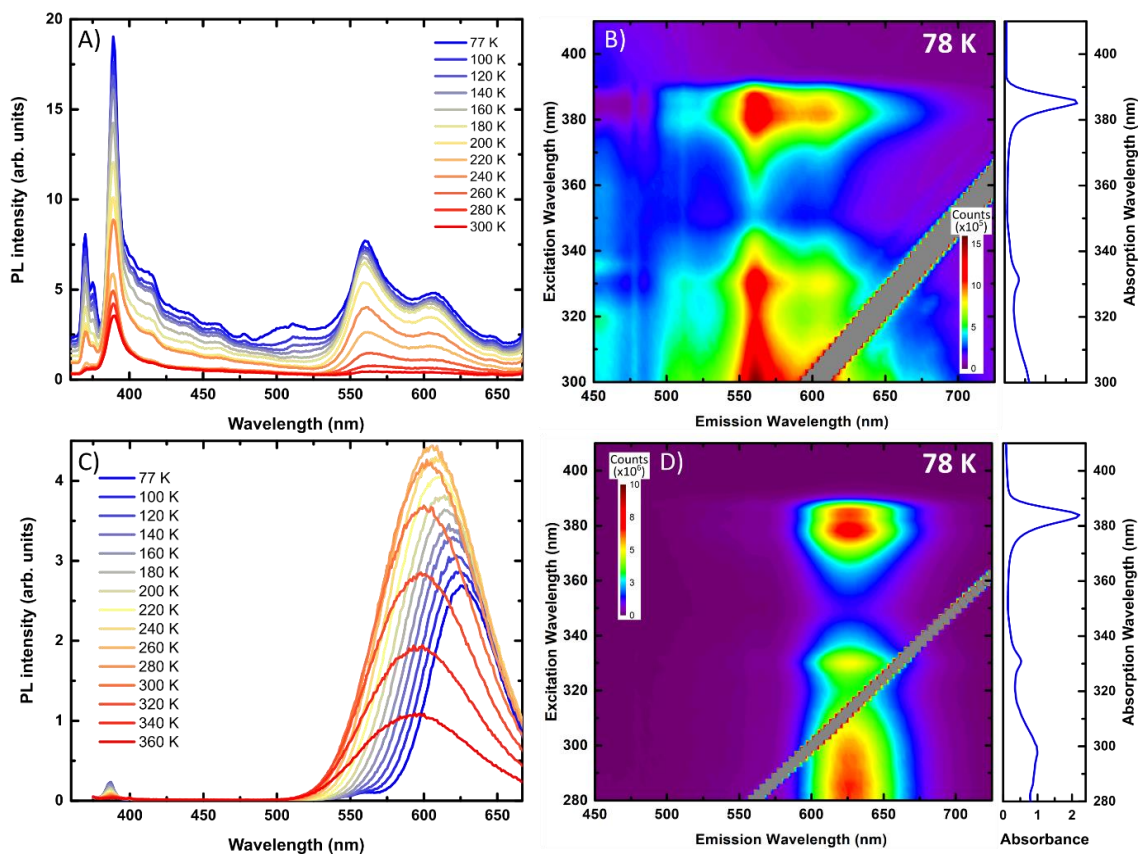

**Figure 3. Temperature-dependent photoluminescence.** NMA$_2$PbBr$_4$ A) temperature dependent PL ($\lambda_{exc}$=340nm) and B) photoexcitation map at 78 K. NMA$_2$PbBr$_4$:Mn(10%) C) temperature dependent PL ($\lambda_{exc}$=365nm) and D) photoexcitation map at 78 K.

Contrary to the perovskite's free excitonic emission, which decreases at higher temperature (Fig. 3C and S8), the temperature dependence of the manganese PL intensity shows a maximum at 260 K, which allows to identify two distinct regimes (Figure 4A). When 77 K < T < 260 K the Mn$^{2+}$ emission increases at higher temperature, likely as a consequence of improved exciton diffusion from the perovskite to the emissive doping centres. Thanks to the favourable alignment of the NMA triplet state (T$_1$), transfer could proceed via two channels schematically depicted in the energy level diagram of Fig 4B: i) direct transfer from the perovskite's excitons to the recombination centres introduced by the Mn$^{2+}$ guest; ii) cascade transfer from the perovskite's excitons to the NMA triplet state, followed by final transfer to the Mn$^{2+}$ centres. Both mechanisms compete with the exciton trapping at the perovskite defect sites that, in the pristine material, result in PL broadening and promote non-radiative recombination causing severe luminescence quenching. In the doped system, funnelling to Mn$^{2+}$ introduces an additional channel which allows to overcome the detrimental trapping at perovskite defects, favouring radiative recombination at the doping sites and finally enhancing the PLQY. When T > 260 K, the PL intensity quickly drops since de-trapping from Mn$^{2+}$ sites become more favourable (likely involving back-transfer to the NMA triplet state, Fig. 4B), jointly with increased exciton thermal quenching up to 360 K (Fig. 4A). We also performed time-resolved photoluminescence measurements (TRPL) under photoexcitation at 355 nm (3.49 eV) at room temperature. The emission from (NMA)$_2$PbBr$_4$ (Fig. 4C) decays in few hundreds of µs, a typical time range for triplet excitation.[53] Upon manganese addition (Fig. 4D), the PL spectrum consists in a main band at 620 nm with a lifetime extended to thousands of µs, characteristic of the formally spin- and parity-forbidden $^4T_1 \rightarrow\ ^6A_1$ transition of $Mn^{2+}$ which slows down the radiative recombination.[51,59] The power dependence of the manganese luminescence shows a maximum of relative PLQY at the excitation density $10^{17}$ $cm^{-3}$, which might stem from exciton-exciton annihilation processes occurring at higher excitations even before reaching the saturation of the $Mn^{2+}$ emissive centres.

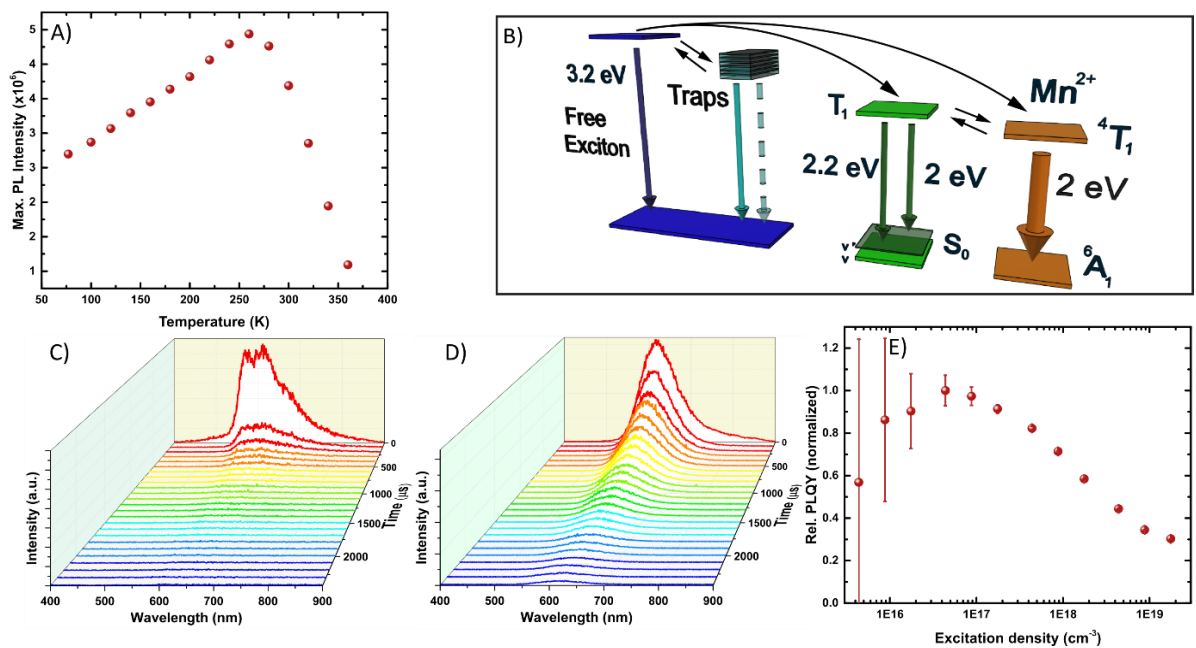

**Figure 4. Transfer processes in the Mn-doped perovskite.** A) Temperature dependence of the maximum photoluminescence intensity of $NMA_2PbBr_4$:Mn(10%); B) schematic representation of the energy level diagram in the Mn-doped system. The blue levels represent the perovskite's excitonic transition. The light-blue levels represent permanent traps with wide energy distribution, with the two arrows showing radiative (continuous line) and non-radiative (dashed line) recombination processes. $T_1$ and $S_0$ (green levels) are respectively the triplet and ground singlet state of NMA. The radiative transitions from $T_1$ to the two vibrational levels of $S_0$ (v and v') are represented, with energy 2.2 eV (560 nm) and 2.0 eV (605 nm). The orange levels represent the $^4T_1 \rightarrow\ ^6A_1$ transition, based on the 3d orbitals of $Mn^{2+}$; C) Time-resolved photoluminescence spectra of C) $NMA_2PbBr_4$ and D) $NMA_2PbBr_4$:Mn(10%). The time scales are in µs; E) relative photoluminescence quantum yield (PLQY) against excitation density for $NMA_2PbBr_4$:Mn(10%).

To test the electroluminescence properties of the material, we integrated $(NMA)_2PbBr_4$:Mn (10%) as emitter in a perovskite light-emitting diode (PeLED) architecture. The structure of the device was ITO/PEDOT:PSS/PolyTPD/$NMA_2PbBr_4$:Mn(10%)/TPBi/Ba/Al, where PEDOT:PSS is a hole injecting layer, PolyTPD plays a role of a hole transporting material, while TPBi is an electron transporting material (see Fig. 5A). Due to the low conductivity of the wide band gap lead bromide perovskite, the use of extremely thin perovskite layers was compulsory to allow charge transport and radiative recombination within the perovskite film. Spin coating from low concentration (0.066 M) DMF perovskite solutions on PolyTPD yielded 30 nm thick films made of discontinuous perovskite islands (Fig. 5B,C). Although the imperfect coverage creates extra current-loss channels in the device, the presence of thin perovskite islands allowed us to probe radiative recombination of injected charges within the perovskite domains at room temperature (Fig. 5C). The shape of the EL spectrum (Fig. 5D) closely resembles the PL spectrum of the doped material, proving that the emission is mainly originating from manganese centres, although a minor contribution from the NMA triplet emission might also be involved (See Fig. S9). A small red-shift (10 nm) of the EL with respect to the PL is most likely caused by different mechanisms of excitation responsible for both emissions. In PL the whole bulk of the material is excited uniformly, therefore states with a wide energy distribution are populated evenly. On the other hand, in the case of electrical excitation charge carriers preferentially populate states with the lowest energy, hence emission is red-shifted with respect to PL. Finally note that in Fig. S9, the EQE of a non-doped PeLED with the same architecture as the Mn-doped one is also presented. The diode employing $NMA_2PbBr_4$ without doping is almost one order of magnitude less efficient; this feature reflects well the measured PLQY values of both materials and supports the doping strategy as a possible way for luminescence tuning and efficiency improvement of 2D perovskites.

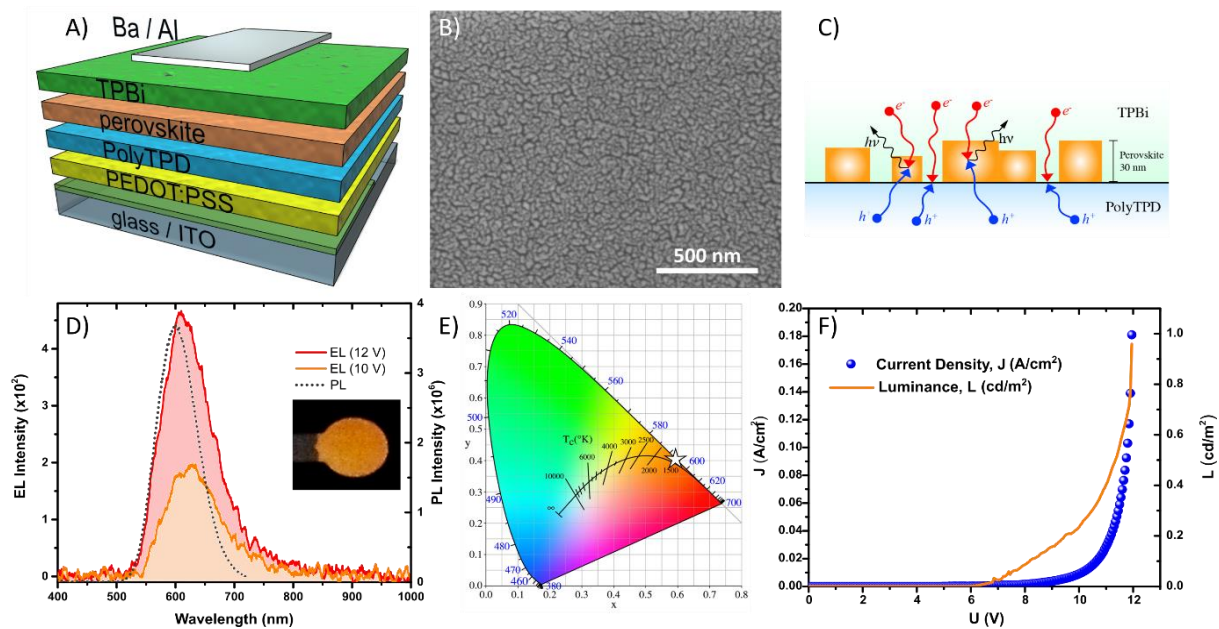

**Figure 5. Perovskite integration in light-emitting diode (LED).** A) LED architecture; B) SEM image of the $NMA_2PbBr_4$:Mn(10%) film morphology when deposited on top of PolyTPD; C) schematic representation of the island-like morphology of the perovskite film, with thickness ~30nm. The arrows represent electron ($e^-$) and hole ($h^+$) parasitic recombination at the PolyTPD/TPBi interface, as well as their radiative recombination within the perovskite domains; D) electroluminescence (EL) spectra measured in a $NMA_2PbBr_4$:Mn(10%) based LED, with the inset showing the orange emission from the device; E) CIE coordinates (0.59, 0.41) of the orange EL; F) Current density-voltage-luminance plot of the $NMA_2PbBr_4$:Mn(10%) LED.

CIE coordinates of the EL spectrum are x = 0.59, y = 0.41 (Fig. 5E), which correspond to orange colour, as confirmed by a photograph of the working device in the inset of Fig. 5D. Current density-voltage-luminance plot of the device is presented in Fig. 5F. The current density curve has the shape typical for a diode, while the luminance curve does not follow the same trend. Such discrepancy is probably caused by the poor conductive properties of the 2D perovskite and the not optimal electrical conditions for efficient emission from the material, which ultimately lead to low luminance values at relatively high current densities.

In summary, we have investigated the effects of Mn-doping in $NMA_2PbX_4$, where NMA=1-naphthylmethylammonium and X=Br, I. Even though transfer from the perovskite inorganic layers can occur to both NMA and $Mn^{2+}$, we showed that only after manganese introduction a remarkable improvement of luminescence efficiency is achieved, reaching PLQY > 22% in doped films. This indicates that the insertion of the metallic heteroatom into the perovskite framework is fundamental to guarantee highly efficient energy transfer to the doping centres overcoming trap-mediated non-radiative recombination pathways. For the first time, we have shown electroluminescence from the guest metallic ions ($Mn^{2+}$) in a perovskite host structure achieved in light-emitting diodes with room-temperature operability, with improved properties compared to the pristine perovskite. This proof-of-concept demonstration shows the effectiveness of inorganic doping as alternative way to tune and enhance the optoelectronic properties of layered perovskites, paving the way for the study of more efficient materials. Our work motivates the study of different combination of templating organic cations and metallic ion guests (e.g., $Cu^+$ and lanthanides) and prompts the deeper investigation of the transfer dynamics in such complex systems. We foresee that the development of new deposition methods to obtain uniform ultra-thin 2D perovskite films, and the use of multidimensional perovskites as host structures will considerably improve the charge transport as well as the device efficiency.

**EXPERIMENTAL PROCEDURES**

The experimental procedures are included in the Supplemental Information.

**SUPPLEMENTAL INFORMATION**

Supplemental Information includes Supplemental Experimental Procedures and 9 figures.

**AUTHORS CONTRIBUTIONS**

D.C. and W.M. contributed equally to this work. D. C. and A.P. conceived the idea for the manuscript and designed the experiments. D. C. synthetized and characterized the materials. W. M. fabricated and characterized the PeLEDs. S. N., T. B. and G. F. performed spectroscopic characterizations. A. P. supervised the work. All authors contributed to the data analysis, discussed the results and contributed to the writing of the manuscript.

**ACKNOWLEDGEMENTS**

This work has been in part funded by the ERC project SOPHY under grant agreement N 771528. We thank Alex J. Barker for technical support and Stefano Perissinotto for assistance in the SEM characterization.

# Supplemental Information

## Mn(II)-doped 2D perovskite for light emitting devices

Daniele Cortecchia, Wojciech Mróz, Stefanie Neutzner, Tetiana Borzda, Giulia Folpini
Annamaria Petrozza

*Istituto Italiano di Tecnologia, Milan, 20133, Italy*

## 1 Experimental methods

### 1.1 Materials.

1-Naphtylmethylamine (97%, Sigma Aldrich), lead bromide $PbBr_2$ (TCI), lead iodide $PbI_2$ (TCI), dimethylformamide DMF (anhydrous, Sigma Aldrich), hydriodic acid HI (57% in water, stabilized), hydrobromic acid HBr (48% in water, Sigma Aldrich), tetrahydrofuran THF (99.9%, Sigma Aldrich), chlorobenzene CB (anhydrous, Sigma Aldrich), dichloromethane DCM (anhydrous, Sigma Aldrich), poly(3,4-ethylenedioxythiophene) polystyrene sulfonate PEDOT:PSS (Heraeus Clevios P VP AI 4083), Poly[N,N'-bis(4-butylphenyl)-N,N'-bis(phenyl)-benzidine] PolyTPD (Ossila), 1,3,5-Tris(1-phenyl-1-H-benzimidazol-2-yl)benzene TPBi (Ossila), E132 PV & LED Encapsulation Epoxy (Ossila).

### 1.2 Synthesis of 1-naphtylmethylammonium bromide (NMA)Br and iodide (NMA)I.

1-Naphtylmethylamine (1.5 ml, 0.01 mmol) was dissolved in 40 ml of tetrahydrofuran (THF) and 3.3 ml of HBr 48% (3 equivalents) were added dropwise to the solution kept in an ice bath under vigorous magnetic stirring. The reaction was stopped after 3h, and the product was precipitated from THF by adding dichloromethane (DCM). This washing procedure was repeated 3 times, and the final product (NMA)Br was collected in form of a white powder by drying it under vacuum at 60 °C in a rotary evaporator. A similar synthetic procedure was adopted for (NMA)I, by substituting HBr with HI (57% water solution, stabilized).

### 1.3 Perovskite synthesis.

For the synthesis of the perovskite $NMA_2PbBr_4$, (NMA)Br and $PbBr_2$ were mixed in 2:1 molar ratio in dimethylformamide (DMF) in the desired concentration (typically 0.25M for thin films used for optical characterization). In the case of the doped perovskite $NMA_2PbBr_4$:Mn(x%), $MnBr_2$ was added to the precursor solution in the desired stoichiometric ratio, with the manganese content varying from 2% to 30% of the lead content. For example, in the case of $NMA_2PbBr_4$:Mn(10%), 0.25M solutions were prepared by mixing 59.5 mg (0.25 mmol) of (NMA)Br, 45.9 mg (0.125 mmol) of $PbBr_2$ and 2.68 mg (0.0125 mmol) $MnBr_2$ in 0.5 ml of DMF. $NMA_2PbI_4$:Mn(10%) were synthetized in a similar way by mixing (NMA)I, $PbI_2$ and $MnBr_2$. The mixture was heated at 100 °C for 2h, and then spin coated on quartz substrates at 4000 rpm for 30 s, and the resulting film was annealed at 100 °C for 15 min. In the case of $NMA_2PbBr_4$ powders, the precursor DMF solution was dried under vacuum at 80 °C in a rotary evaporator, and finally annealed at 100°C for 4 h in vacuum oven.

### 1.4 Structural and morphological characterization.

Perovskite powders and thin films were characterized through X-ray diffraction (XRD) using a BRUKER D8 ADVANCE with Bragg-Brentano geometry, Cu Kα radiation (λ = 1.54056 Å), step increment of 0.02° and 1 s of acquisition time. Films used for XRD analysis were spin coated from 0.25M solutions in DMF. The surface profiler Dektak 150 (Veeco) was used to measure the film thickness. The scanning electron microscope (SEM) MIRA3, TESCAN was used to characterize the film morphology. Before characterization, 7 nm of gold were evaporated on top of the thin film to avoid charging effects during the image acquisition.

### 1.5 Spectroscopic characterization.

Steady state absorption spectra were measured on perovskite thin films deposited on quartz using a UV/VIS/NIR spectrophotometer Lambda 1050, Perkin Elmer equipped with integrating sphere (module 150mm InGaAs Int. Sphere). Photoluminescence emission spectra were taken with a NanoLog (Horiba Jobin Yvon), while the absolute photoluminescence quantum yield (PLQY) was measured using an integrating sphere (Quanta-φ) in combination with the NanoLog spectrofluorometer. Low temperature measurements were performed in an Oxford Instruments cold finger cryostat MicrostatHEcooled with liquid nitrogen. Time-resolved photoluminescence measurements were performed by exciting the sample with the 3rd harmonic (355 nm) from a Nd:YAG Picolo-AOT laser with a pulse length of approximately 1000 ps and pulse fluence of 120 µJ/cm$^2$. The PL was detected with an Andor iStar 320T ICCD camera coupled to a Shamrock 303i spectrograph using a temporal step size of 100 µs and a spectral resolution of 0.2 nm. All measurements were performed in vacuum. The power dependent photoluminescence measurements were taken on thin films, deposited on sapphire to guarantee good thermal contact. The same light source as for the time-resolved photoluminescence measurements was used (Picolo-AOT) with 355 nm excitation wavelength. During the measurements, the sample was kept in the cold finger cryostat at 260 K to prevent sample degradation.

### 1.6 Fabrication of light-emitting diodes (LEDs).

Glass substrates with indium tin oxide (ITO) were etched with Zn powder and 2M HCl. The etched substrates were cleaned ultrasonically in distilled water, acetone and isopropanol and subsequently O$_2$ plasma treated for 10 min. Afterwards, poly(3,4-ethylenedioxythiophene) polystyrene sulfonate (Heraeus Clevios P VP AI 4083) (PEDOT:PSS) was spin-coated through PVDF filter (pore size 0.22 µm) on substrates rotating 4000rpm, followed by thermal annealing at 110 °C for 10 min in air. The deposited PEDOT:PSS layers had thicknesses around 50 nm. Then substrates were transferred to a N$_2$ filled glovebox where the rest of devices' construction processing followed. A chlorobenzene solution of Poly[N,N'-bis(4-butylphenyl)-N,N'-bis(phenyl)-benzidine] (PolyTPD) with concentration 15 mg/ml was spin-coated on PEDOT:PSS with 4000 rpm, resulting in a film of 25 nm thickness. The film was annealed at 100 °C for 10 min and subsequently O$_2$ plasma treatment was applied for 5 s (outside the glovebox) to improve wettability of the PolyTPD film. On such substrates a hot (100 °C) 0.066 M dimethyl formaldehyde (DMF) solution of (NMA)$_2$PbBr$_4$:Mn(10%) was spin-coated with 6000 rpm and devices were annealed at 100 °C for 15 min, achieving a final film thickness of about 30 nm. Afterwards, a 35 nm thick film of 1,3,5-Tris(1-phenyl-1-H-benzimidazol-2-yl)benzene (TPBi) was thermally evaporated on the samples at a pressure $1*10^{-6}$ mbar. Subsequently, the samples were transferred to a metal evaporator, where 7 nm of barium and 100 nm of aluminium were evaporated through a shadow mask, creating in this way a cathode. The active area of the devices was 8.3 mm$^2$. The light-emitting diodes were encapsulated by placing a single drop of a resin (E132 PV & LED Encapsulation Epoxy, Ossila) on the active area, covering it with a microscope glass slide and UV curing for 20 min.

### *1.7 Device characterization.*

Current–voltage (I–V) device characterization was performed with a Keithley 2602 source meter combined with a photodiode with known responsivity (OSI Optoelectronics, UV-100DQC). Measurements were realized by applying bias to the diode while the photodiode in contact with the device registered the emitted light. Electroluminescence was collected by an optical fiber and recorded by an Ocean Optics Maya 2000 Pro spectrometer. Efficiencies were calculated assuming a Lambertian distribution of emission.

### *2 Supporting figures.*

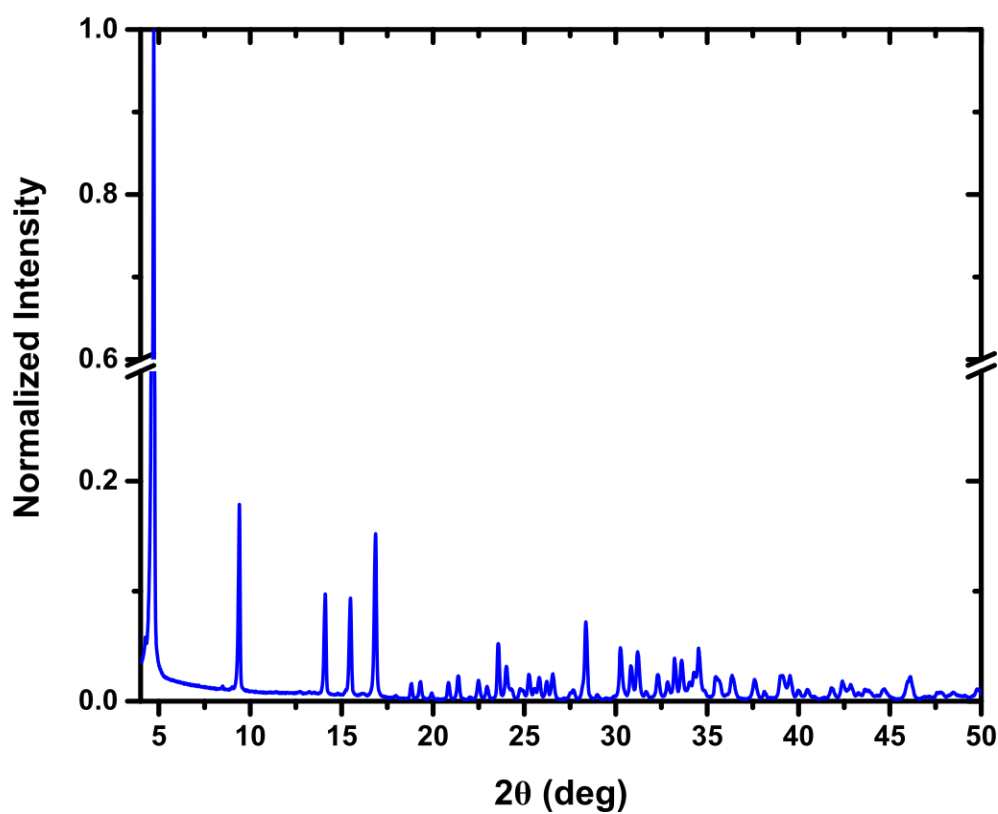

**Figure S6.** (NMA)$_2$PbBr$_4$ powder XRD.

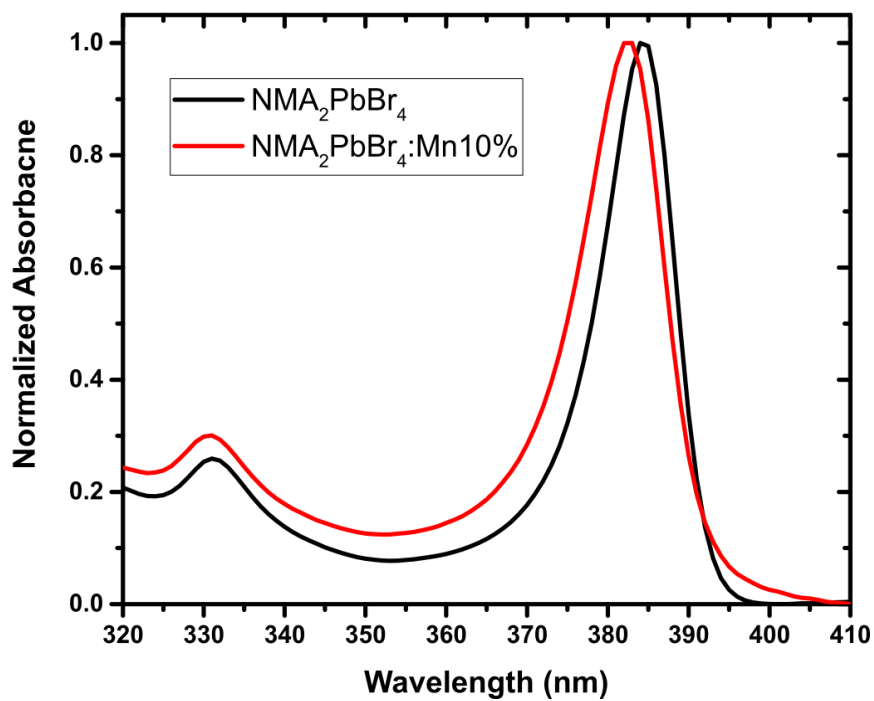

**Figure S7.** NMA$_2$PbBr$_4$ (black line) and NMA$_2$PbBr$_4$:Mn(10%) room temperature absorption spectra, showing a ~3 nm blue-shift and ~3 nm broadening of the excitonic absorption in the doped perovskite film.

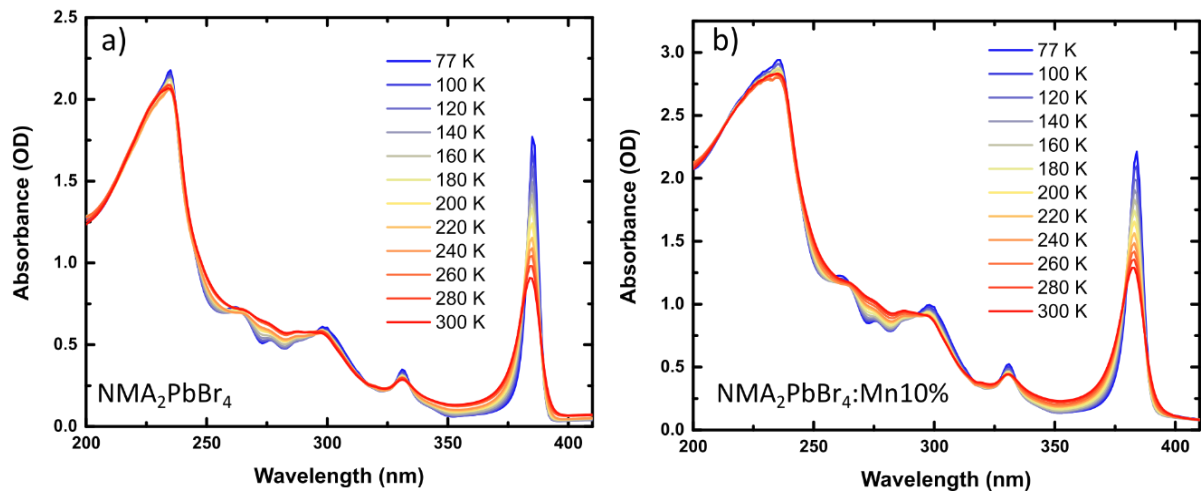

**Figure S3.** Temperature dependent absorption of a) NMA$_2$PbBr$_4$ and b) NMA$_2$PbBr$_4$:Mn(10%) thin films.

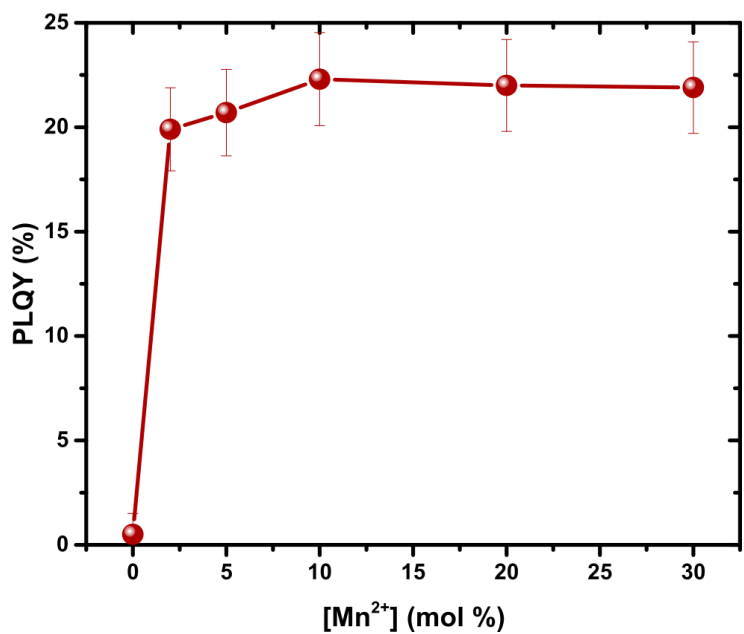

**Figure S4.** Dependency of the photoluminescence quantum yield (PLQY) of $NMA_2PbBr_4:Mn(x)$ on the $Mn^{2+}$ content of the starting reaction solution. This is expressed as x=0-30% mol, indicating the $Mn^{2+}$ molar content respect to the lead amount. The PLQYs of both $NMA_2PbBr_4$ and $NMA_2PbI_4$ pristine perovskites were estimated to be < 1%, in a range comparable to or below the instrumental resolution.

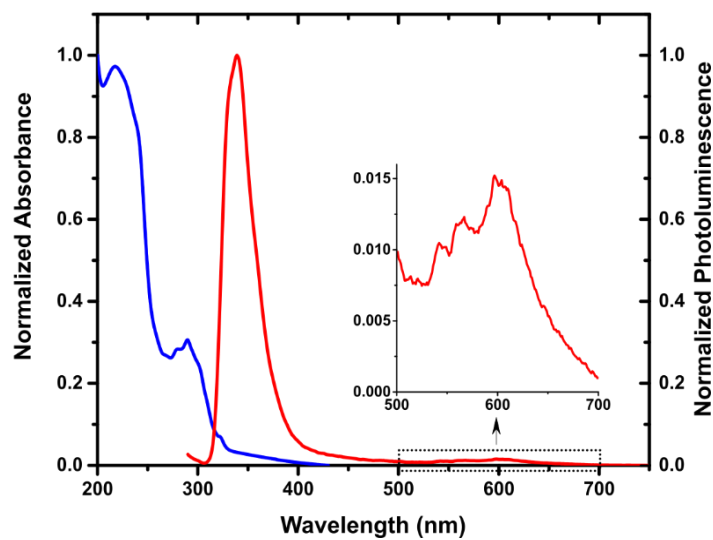

**Figure S5.** Absorption (blue line) and photoluminescence (red line) of the organic salt (NMA)Br at $\lambda_{exc}$=270 nm.

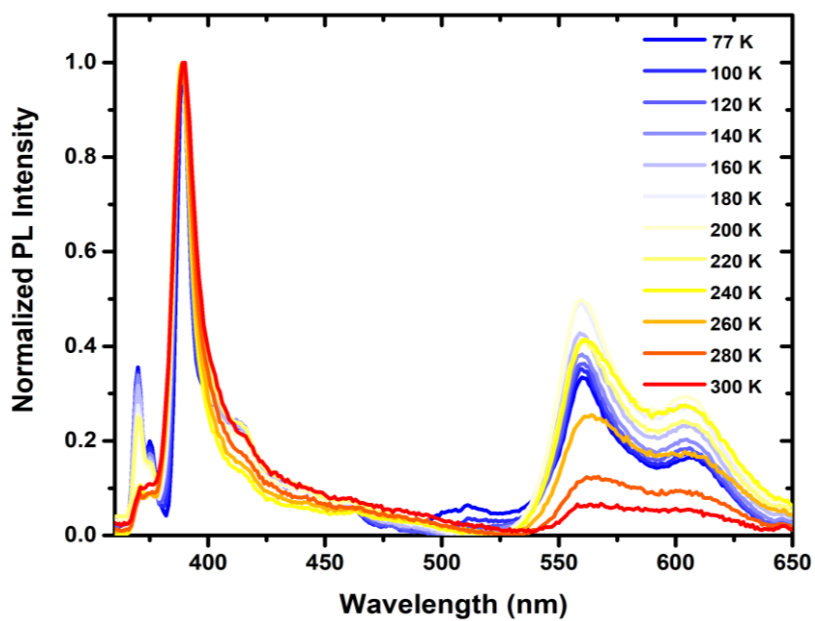

**Figure S6.** Temperature dependent, normalized photoluminescence of NMA$_2$PbBr$_4$. Excitation wavelength $\lambda_{exc}$=340 nm.

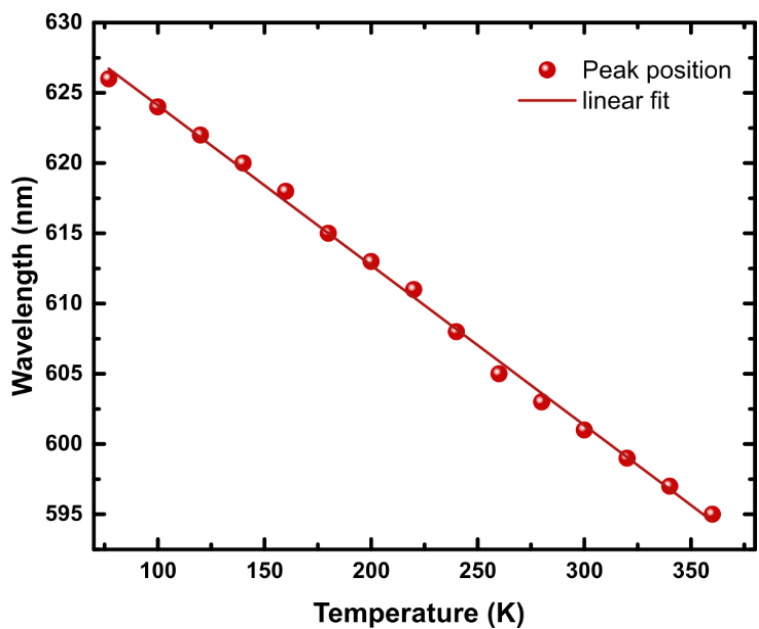

**Figure S7.** Temperature dependent shift of the photoluminescence peak maximum in NMA$_2$PbBr$_4$:Mn(10%).

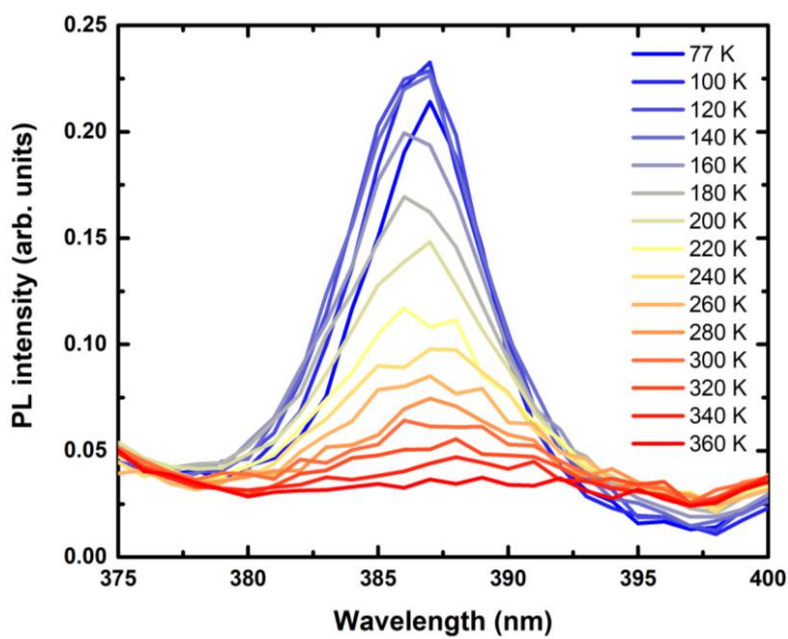

**Figure S8.** Detail of the temperature dependent excitonic luminescence in the doped perovskite NMA$_2$PbBr$_4$:Mn(10%). Excitation wavelength $\lambda_{exc}$=365 nm.

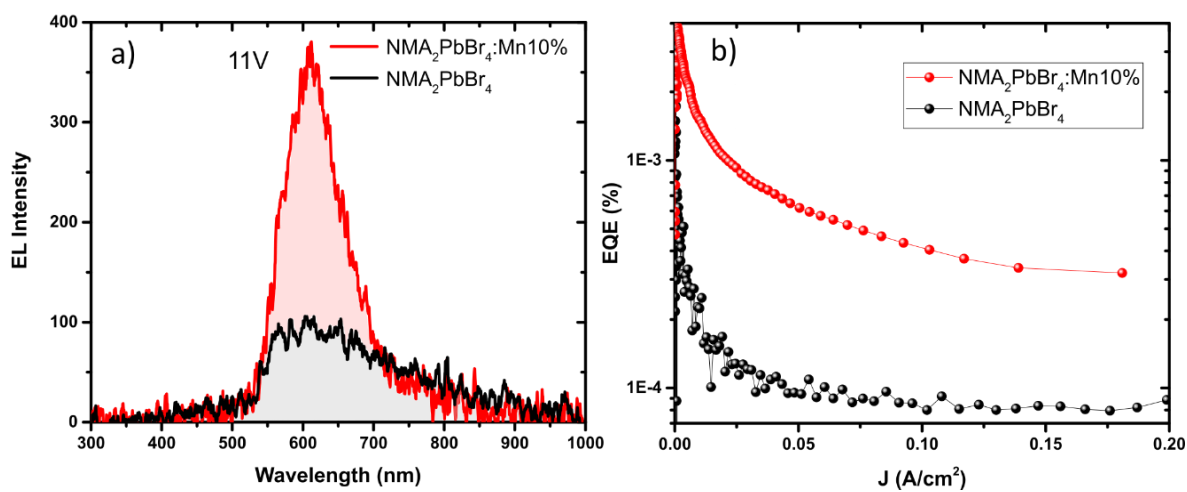

**Figure S9.** a) Comparison of EL spectra of the non-doped (black line) and doped (red line) perovskite LED at 11 V. The PeLED with Mn inclusion has more intense and narrower spectrum in comparison to the non-doped one. These properties confirm the different nature of the emissions: while for the doped perovskite it originates from Mn centres, in the case of the non-doped device it originates from the NMA phosphorescence. b) External quantum efficiency (EQE) comparison between the LED employing NMA$_2$PbBr$_4$ (black line) and NMA$_2$PbBr$_4$:Mn(10%) (red line) as emitting layer.